\documentclass[useAMS,usenatbib,letter,amsmath,fleqn,preprint]{aastex}
\usepackage{graphicx,color}
\usepackage{amssymb,amsmath}
\usepackage{float}
\usepackage{longtable}
\usepackage{braket}
\numberwithin{equation}{section}

\title{Equilibrium rotation states of doubly synchronous binary asteroids}
\author{Oleksiy Golubov\altaffilmark{1,2,3} and Daniel J. Scheeres}
\affil{Department of Aerospace Engineering Sciences, University of Colorado at Boulder \\
429 UCB, Boulder, CO, 80309, USA}

\altaffiltext{1}{Karazin Kharkiv National University, 4 Svobody Sq., Kharkiv, 61022, Ukraine}
\altaffiltext{2}{Institute of Astronomy of V. N. Karazin Kharkiv National University, 35 Sumska Str., Kharkiv, 61022, Ukraine}
\altaffiltext{3}{olexiy.golubov@gmail.com}

\begin{abstract}
A doubly synchronous binary asteroid simultaneously experiences YORP and BYORP,
the former being independent of the radius of the orbit, while the latter linearly dependent on the radius.
In many systems YORP and BYORP can compensate each other at some radius,
causing an equilibrium state, to which other orbits converge in the course of their evolution.
We derive mathematical formalism of such equilibria in terms of dimensionless YORP and BYORP coefficients.
We compute YORP and BYORP coefficients for a set of photometric and radar shape models,
and find equilibria to be relatively common.
\end{abstract}


\begin{document}

\section{Introduction}
A doubly synchronous binary asteroid is subject to the BYORP effect, 
which can either increase or decrease its angular momentum \citep{cuk05}.
Simultaneously both components of the system experience the YORP effect \citep{rubincam00}.
Under some conditions a stable equilibrium berween these two torques is possible,
and it is the subject of this article.

Interaction between YORP and BYORP torque is non-trivial.
While BYORP inputs momentum into orbital motion of the two components \citep{mcmahon10},
YORP alters rotation rates of the two components separately.
The orbital period and the two rotational periods of the components get slightly out of synchronization,
and they turn with respect to each other, until the further offset is stopped gravity torques,
which originate when the binary system gets lopsided.
These gravity torques help to distribute the torqus inside the binary system.
If the exterior torqus are small enough, the system remains tidally locked.
When studying dynamics of such a system, we can just add up the BYORP and the two YORP torques.
Although the torques are acting on different bodies, they are re-distributed in such a way, 
that the system remains synchronized, and rotates as a whole.

\begin{figure}
 \centering
 \includegraphics[width=0.48\textwidth]{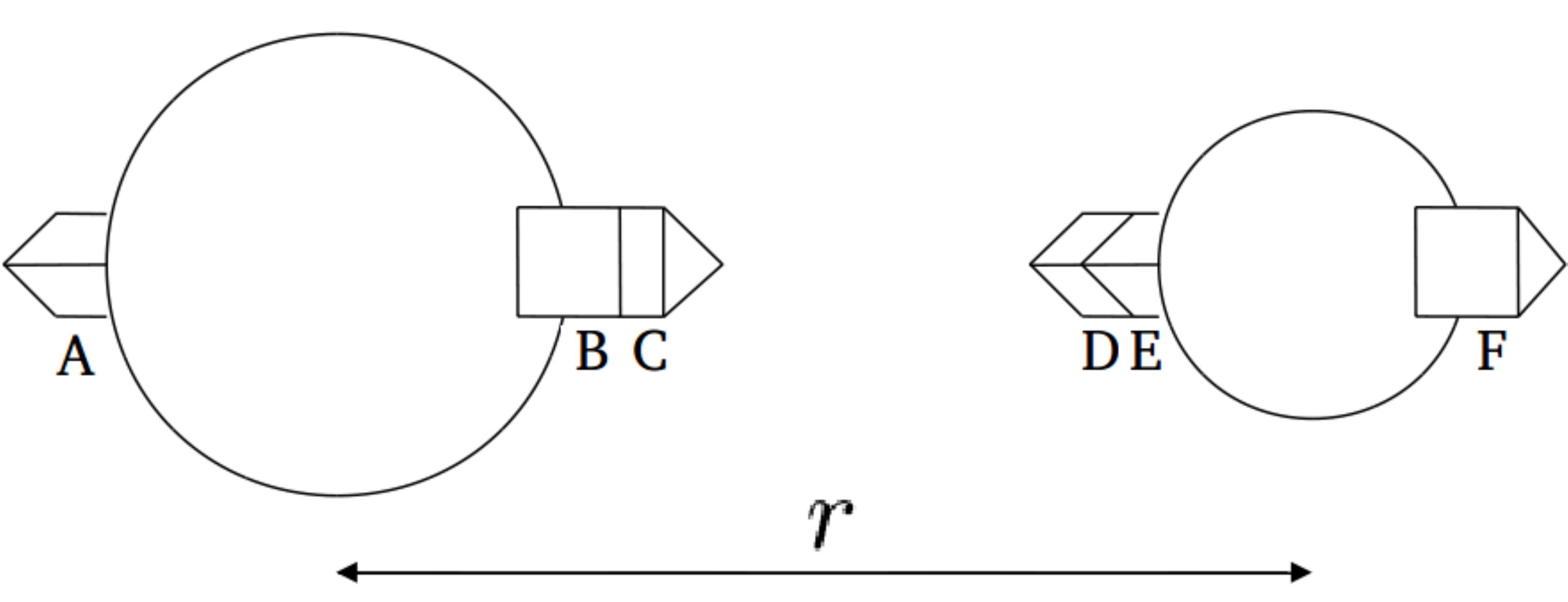}
 \caption{Sketch of a system capable of YORP-BYORP equilibrium.}
 \label{fig-rubincam}
\end{figure}

The idea of an equilibrium between YORP and BYORP is illustrated in Figure \ref{fig-rubincam}.
The binary system is composed of two Rubincam propellers with wedges of slightly different height.
Then wedges A and B create YORP torque on the primary, wedges E and F create YORP torque on the secondary,
and the superstructures C and D create BYORP.
The BYORP torque is negative and linearly increases with the distance between the asteroids.
The YORP torque is positive and independent of the distance between the asteroids.
If the superstructures C and D are small enough, the total torque is positive at a close distance between the components.
So the angular momentum of the system increases, and the two components move further apart.
The distance between the superstructures C and D increases,
and the larger lever arm causes the larger BYORP torque.
It is negative, and when the distance between the asteroids is large enough, BYORP ultimately compensates YORP,
the further outward motion is stopped, and the system arrives at an equilibrium.

This basic idea is treated in the paper more rigorously.
In Section \ref{sec:theory} we present the mathematical formalism, which describes entangled YORP--BYORP evolution.
In Section \ref{sec:applications} we apply the formalism to a set of photometric and radar shape models of asteroids,
and in Section \ref{sec:discussion} discuss implications of such equilibria to the evolution of asteroids.

Throughout this paper we neglect tangential YORP \citep{golubov12} and mutual shadowing of the components,
which could alter the torque acting on the system.
We assume obliquity of the system to be fixed, although both YORP and BYORP have components which alter obliquity.
Thus our statements about stability of the orbit are valid only if the obliquity of the orbit is stable.
We know that in many cases orbits with obliquities $0^\circ$, $90^\circ$ or $180^\circ$ are stable, 
thus will we will concentrate on these cases in Section \ref{sec:applications}.
The orbit of the binary system is assumed circular, and evolution of eccentricity is disregarded.
We know that in many cases zero eccentricity is stable with respect to BYORP \citep{mcmahon10}, 
and in some cases when it is not it is damped by tides \cite{jacobson11}.

\section{Theory}
\label{sec:theory}

The axial component of the YORP torque is expressed by the equation \citep{golubov16,steinberg11}
\begin{equation}
T_z = \frac{\Phi}{c} \oint\limits_S (\mathrm{d}\textbf{S}\times\textbf{r})_z \, p_z \,.
\label{T_z-def}
\end{equation}
Here $\Phi=\Phi_0 A_0^2/(\sqrt{1-e^2}A^2)$ is the time averaged solar energy flux at the asteroid's position,
with $\Phi_0$ being the solar constant at the distance $A_0=1$ AU,
$A$ is the semimajor axis of the asteroid's heliocentric orbit, $e$ is its eccentricity,
$c$ the speed of light, and $p_z$ a function depending on the latitude on the asteroid and the asteroid's obliquity \citep{golubov16}.
The integral is to be taken over the surface of both asteroids comprizing the binary.
This equation neglects shadowing and multiple scattering of light from the primary to the secondary, and vice versa.

We direct the $x$ axis from the primary A to the secondary B,
the $z$ axis towards the angular momentum of the system, and the origin of the coordinate system in the center of mass.
Then radius-vectors of points on the two components are
\begin{align}
\textbf{r}_\mathrm{A} = -\mu\textbf{r}+\textbf{r}'_\mathrm{A} = \left(-\mu r+x_\mathrm{A},y_\mathrm{A},z_\mathrm{A}\right)\,,\nonumber\\
\textbf{r}_\mathrm{B} = (1-\mu)\textbf{r}+\textbf{r}'_\mathrm{B} = \left((1-\mu)r+x_\mathrm{B},y_\mathrm{B},z_\mathrm{B}\right)\,.
\label{radius-vectors}
\end{align}
Here $\mu=M_\mathrm{B}/(M_\mathrm{A}+M_\mathrm{B})$ is the mass fraction of the secondary,
$\textbf{r}_\mathrm{A}$ and $\textbf{r}_\mathrm{B}$ are radius-vectors of points on the two components with respect to the center of mass of the system,
$\textbf{r}'_\mathrm{A}$ and $\textbf{r}'_\mathrm{B}$ are radius-vectors with respect to the centers of mass of the two components,
and $\textbf{r}$ is the vector joining the centers of mass of A and B.

Then we substitute Eqn. (\ref{radius-vectors}) into Eqn. (\ref{T_z-def}) and find
\begin{equation}
T_z = \frac{\Phi}{c}\left(R_\mathrm{A}^3 C_\mathrm{A}+R_\mathrm{B}^3 C_\mathrm{B}-aR_\mathrm{T}\mu R_\mathrm{A}^2 B_\mathrm{A}+aR_\mathrm{T}(1-\mu) R_\mathrm{B}^2 B_\mathrm{B}\right)\,.
\label{T_z1}
\end{equation}
Here $R_\mathrm{A}$ and $R_\mathrm{B}$ are volume-equivalent radii of the primary and the secondary, and 
$R_\mathrm{T}=R_\mathrm{A}+R_\mathrm{B}$ is the total radius.
The density of the primary and the secondary is assumed the same, so that
$R_\mathrm{A}=(1-\mu)^{1/3}R_\mathrm{T}/((1-\mu)^{1/3}+\mu^{1/3})$ and $R_\mathrm{B}=\mu^{1/3}R_\mathrm{T}/((1-\mu)^{1/3}+\mu^{1/3})$.
$a=r/R_\mathrm{T}$ is the distance between the components expressed in terms of the total radius.
The dimensionless YORP and BYORP coefficients are determined as
\begin{align}
C_\mathrm{A} = \oint\limits_{S_\mathrm{A}} \left(\frac{\mathrm{d}\textbf{S}_\mathrm{A}^2}{R_\mathrm{A}^2}\times\frac{\textbf{r}_\mathrm{A}}{R_\mathrm{A}}\right)_z \, p_z\,,\nonumber\\
C_\mathrm{B} = \oint\limits_{S_\mathrm{B}} \left(\frac{\mathrm{d}\textbf{S}_\mathrm{B}^2}{R_\mathrm{B}^2}\times\frac{\textbf{r}_\mathrm{B}}{R_\mathrm{B}}\right)_z \, p_z\,,\nonumber\\
B_\mathrm{A} = \oint\limits_{S_\mathrm{A}} \left(\frac{\mathrm{d}\textbf{S}_\mathrm{A}^2}{R_\mathrm{A}^2}\times\textbf{e}_x\right)_z \, p_z\,,\nonumber\\
B_\mathrm{B} = \oint\limits_{S_\mathrm{B}} \left(\frac{\mathrm{d}\textbf{S}_\mathrm{B}^2}{R_\mathrm{B}^2}\times\textbf{e}_x\right)_z \, p_z\,.
\label{YORP-coefficients}
\end{align}
Here $\textbf{e}_x$ is the basis vector in the $x$ direction.
The former two coefficients are the same as dimensionless YORP $\tau_z$ used by \cite{golubov12},
while the latter two are similar to BYORP coefficients used by \cite{jacobson11}.
All the four coefficients are independent of the size of an asteroid, and only depend on its shape.
In general, the coefficients are large only for asymmetric asteroids.

It is convenient to introduce the dimensionless total YORP and BYORP torque, determined as
\begin{equation}
C=\frac{(1-\mu)C_\mathrm{A}+\mu C_\mathrm{B}}{\left((1-\mu)^{1/3}+\mu^{1/3}\right)^3}\,,
\label{C}
\end{equation}
\begin{equation}
B=\frac{a\mu^{2/3}(1-\mu)^{2/3}}{\left((1-\mu)^{1/3}+\mu^{1/3}\right)^2}\left(-\mu^{1/3}B_\mathrm{A}+(1-\mu)^{1/3}B_\mathrm{B}\right)\,.
\label{B}
\end{equation}
Then the total torque experienced by the asteroid can be expressed in a simpler form,
\begin{equation}
T_z = \frac{\Phi}{c}R_\mathrm{T}^3(C+aB)\,.
\label{T_z}
\end{equation}

\begin{figure}
 \centering
 \includegraphics[width=0.48\textwidth]{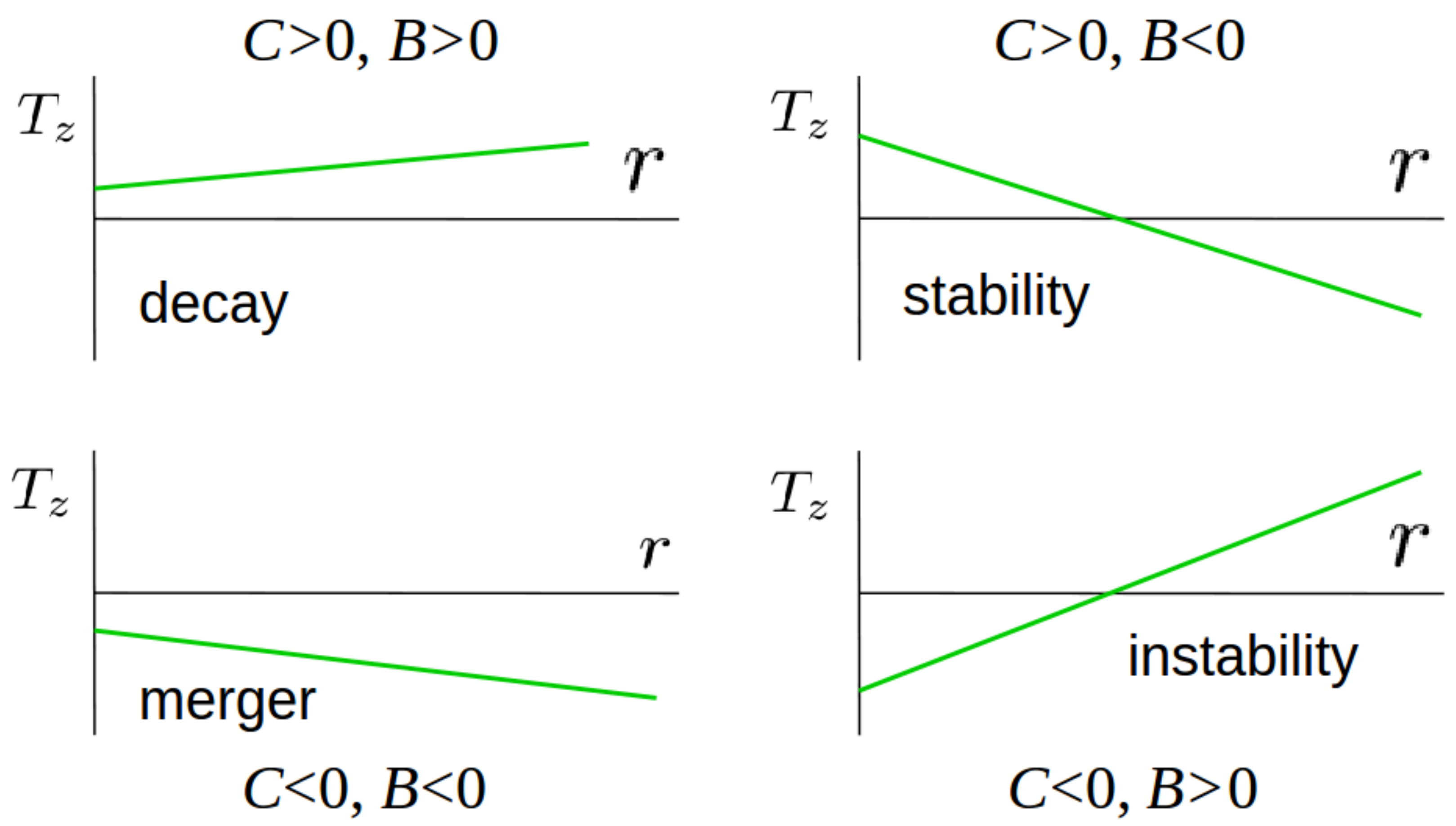}
 \caption{Possible dependencies of $T_z$ on the distance between the components.}
 \label{fig-signs}
\end{figure}

From Eqn. (\ref{T_z}) we see, that the dependence between $T_z$ and $a$ is linear.
Depending on the signs of the coefficients $C$ and $B$, 
one of the four cases presented in Figure \ref{fig-signs} will occur.
Assuming that the system rotates in the positive direction,
in the upper left case the system will decay,
in the lower left case it will merge,
in the upper right case it will reach a stable equilibrium,
while in the lower right case the equilibrium is unstable and the system will either merge or decay depending on the initial conditions.

The upper right case in Figure \ref{fig-signs} is somewhat similar to the stable equilibrium described by \cite{jacobson11},
although the equilibrium is attained not between BYORP and YORP, but between BYORP and tides.
In fact, the Jacobson \& Scheeres equilibrium is stable only from the point of view of the secondary,
while the primary's rotation state continues evolving due to tidal effects,
causing its rotation rate to slow.
In contrast, the equilibrium between YORP and BYORP described here totally stops the evolution of the rotational state.

When the equilibrium is present, it can be determined from Eqn. (\ref{T_z}) by equating $T_z$ to 0.
Thus for the radius of the equilibrium orbit we get
\begin{equation}
a_0 = \frac{|C|}{B}\,.
\label{a_0}
\end{equation}

\section{Applications}
\label{sec:applications}

\begin{figure}
 \centering
 \includegraphics[width=0.48\textwidth]{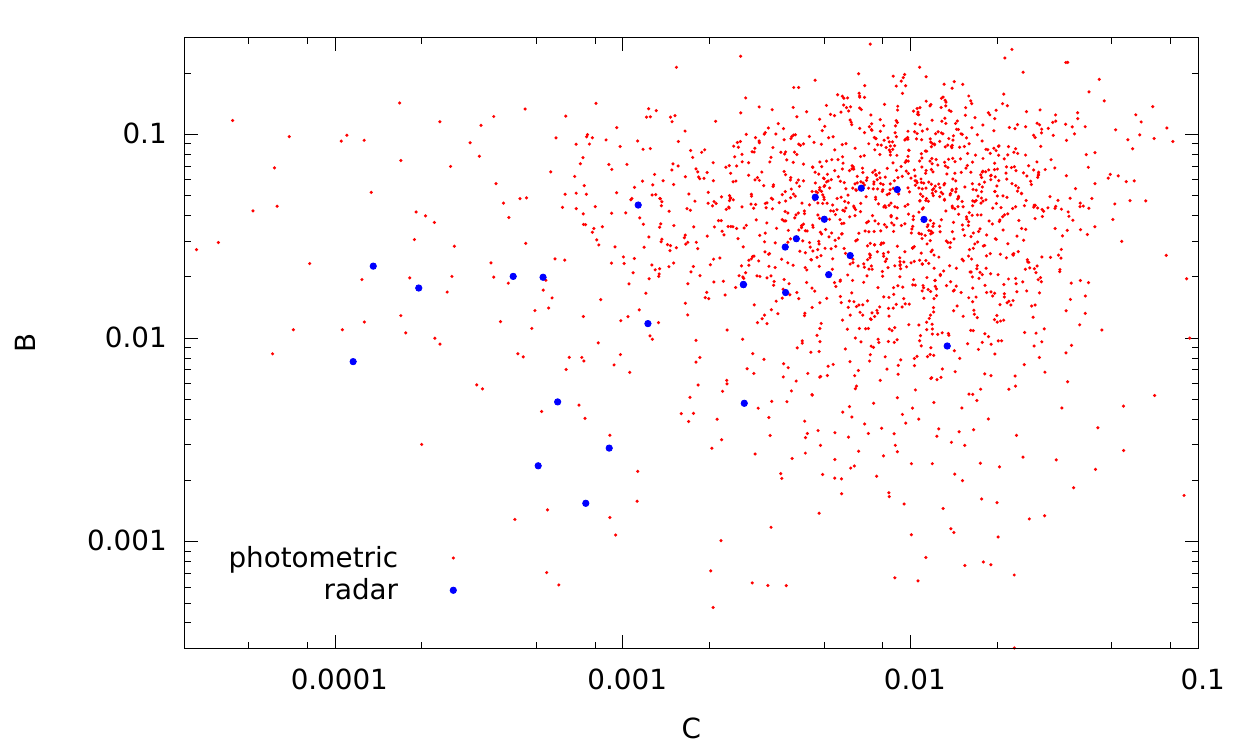}
 \includegraphics[width=0.48\textwidth]{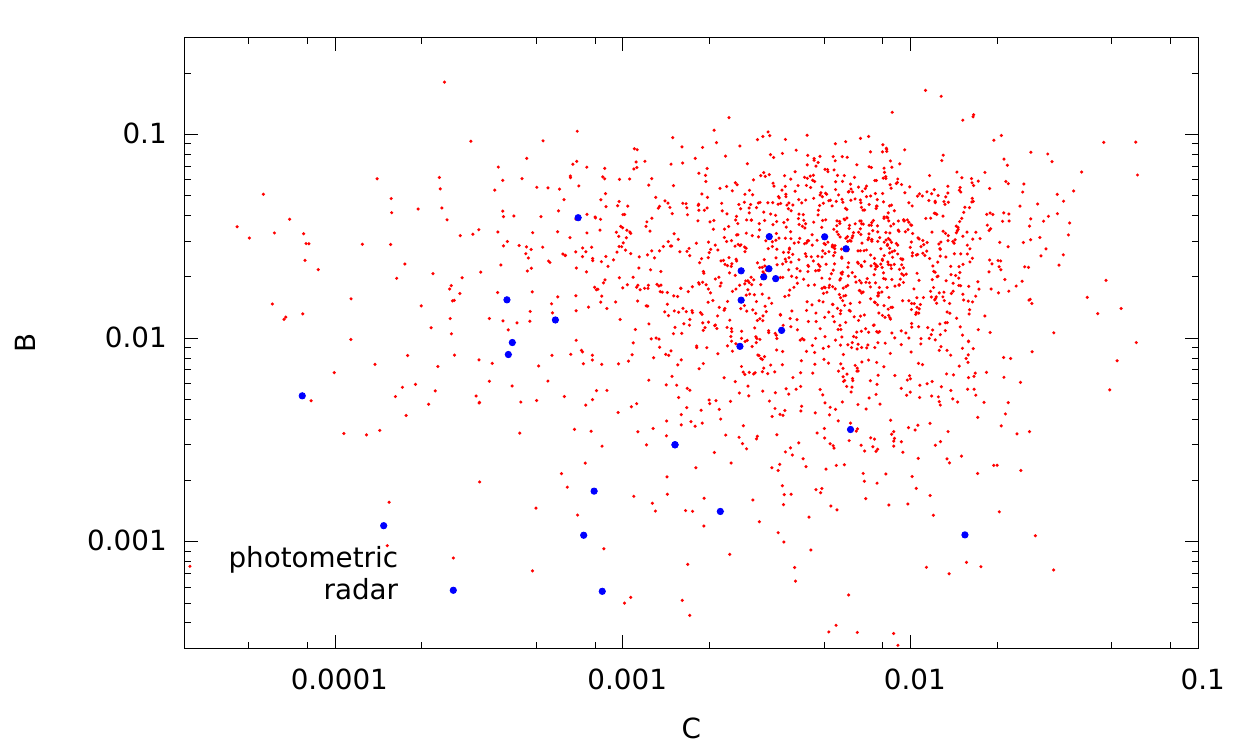}
 \caption{YORP coefficients $C$ and BYORP coefficients $B$ for the sets of photometric and radar models.
 Obliquity is 0 in the left-hand panel and $90^\circ$ in the right-hand panel.}
 \label{fig-coefficients}
\end{figure}

To undertand which YORP and BYORP coefficients are realistic,
we consider a set of photometric and radar shape models,
and compute their YORP and BYORP coefficients using Eq. (\ref{YORP-coefficients}).
The set of photometric shape models contains 1593 shapes of 910 different asteoids from DAMIT database \citep{damit}.
(Some asteroids have two different shape models, and we treat the different models independently.)
The set of radar shapes includes 26 shape models from \cite{radar}.
We plot the resulting coefficients $C$ and $B$ for obliquities $\epsilon=0$ and $\epsilon=90^\circ$ in Figure \ref{fig-coefficients}.

Both $C$ and $B$ have nearly equal probabilities of being positive and negative, 
but for Figure \ref{fig-coefficients} we take their absolute values.
The sign of $C$ is altered if the asteroid spins in the opposite direction,
and the sign of $B$ is altered by a 180$^\circ$ flip of the asteroid around its rotation axis.
So we assume different signs of $C$ and $B$ to be equally probable,actually treating each shape model as four different shape models,
with $\pm C$ and $\pm B$.

\begin{figure}
 \centering
 \includegraphics[width=0.48\textwidth]{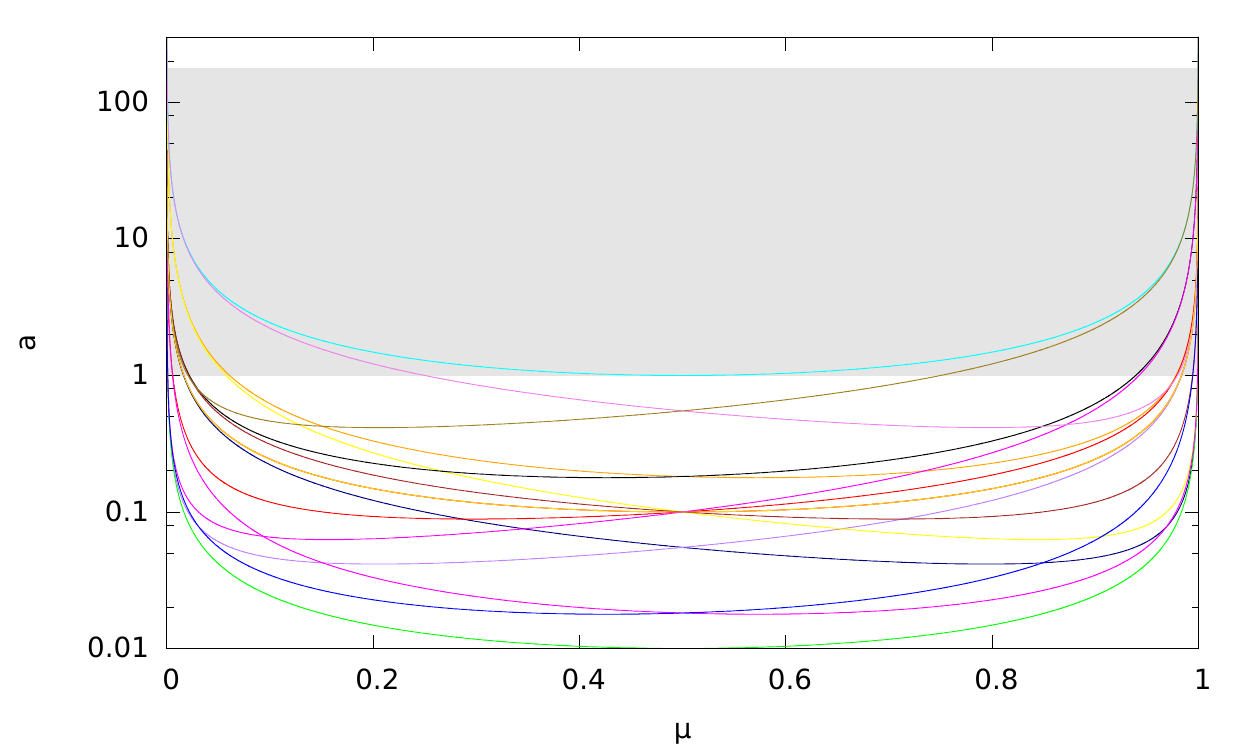}
 \includegraphics[width=0.48\textwidth]{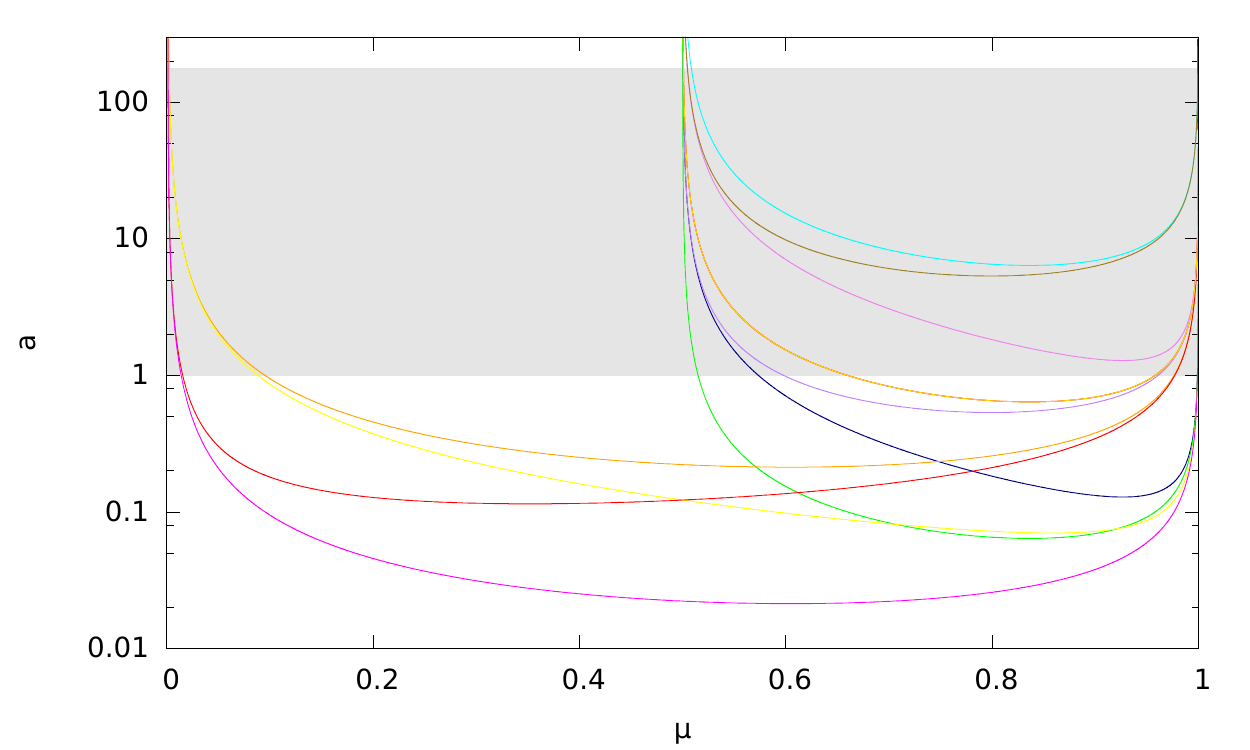}
 \includegraphics[width=0.48\textwidth]{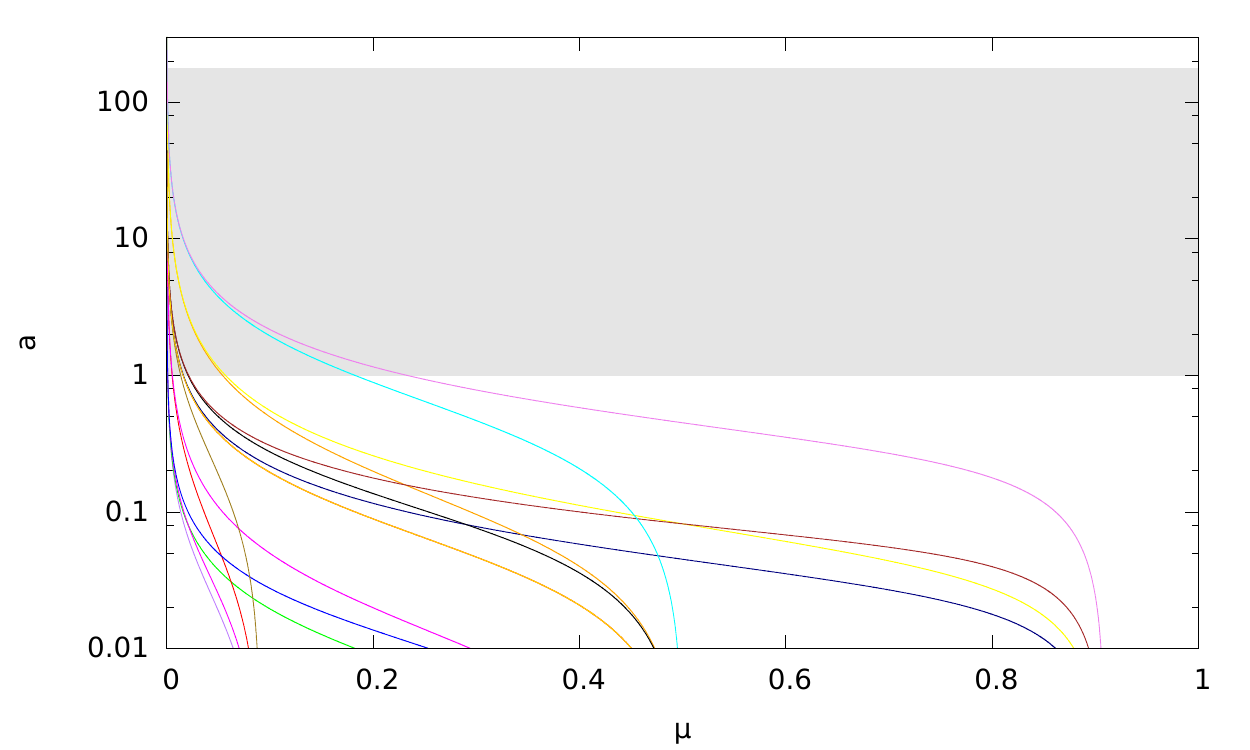}
 \includegraphics[width=0.48\textwidth]{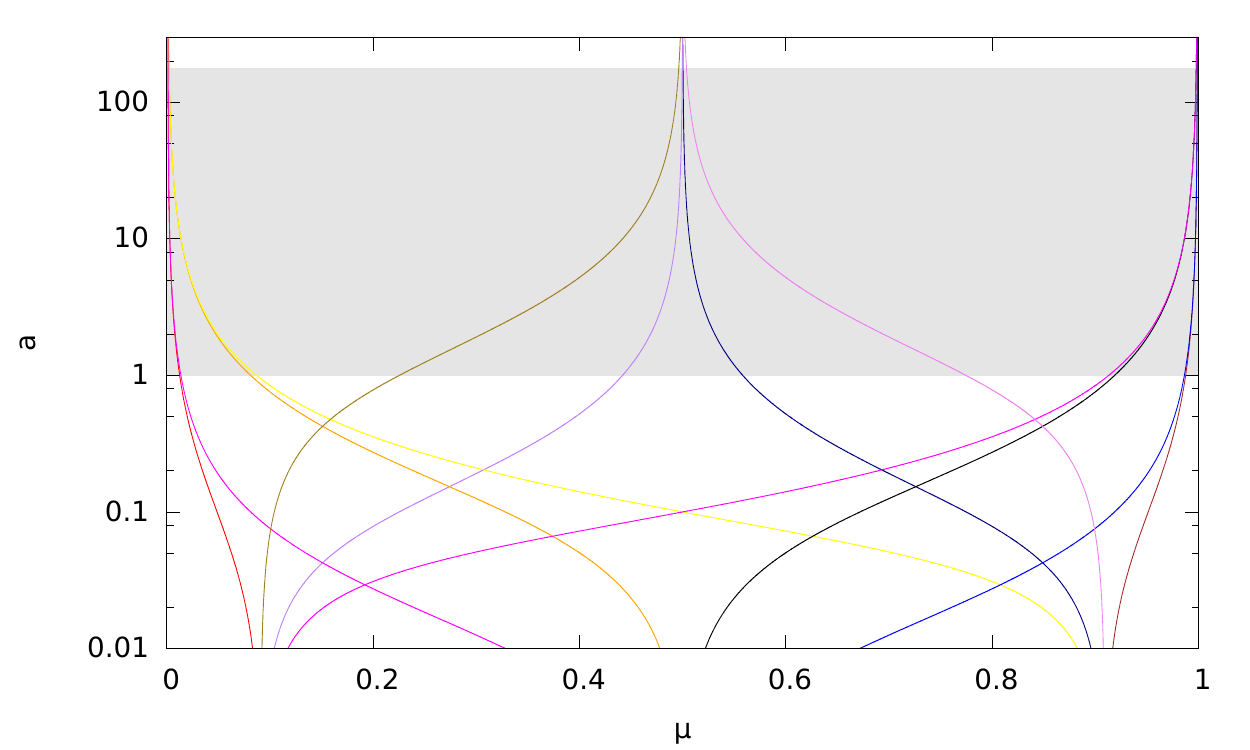}
 \caption{Dependence of the equilibrium distance between the components $a$ on the mass ratio $q$.}
 \label{fig-plots}
\end{figure}

From Figure \ref{fig-coefficients} we see, that values $C_1=0.001$ and $C_2=0.01$, $B_1=0.01$ and $B_2=0.1$ are typical,
so we use them to make the following set of plots.
We fix $C_\mathrm{A}$, $C_\mathrm{B}$, $B_\mathrm{A}$, and $B_\mathrm{B}$ 
at some values among $\pm C_1$, $\pm C_2$, $\pm B_1$, and $\pm B_2$.
We allow different size ratios, thus treating $\mu$ as a parameter.
In Figure \ref{fig-plots} we plot the equilibrium radius $a$ for each size ratio $\mu$.
Different panels of the figure correspond to different signs of YORP and BYORP coefficients.

The shadowed area marks the distances $a$, which not only formally turn Eqn (\ref{T_z}) into 0,
but are also physically feasible.
The smallest possible distance is limited by the two components touching each other,
\begin{equation}
a_\mathrm{min} = 1\,.
\label{a_min}
\end{equation}
The biggest piossible distance is determined by the Hill limit,
\begin{equation}
a_\mathrm{max} = \sqrt[3]{\frac{\rho}{3\rho_\odot}}\frac{A_\mathrm{min}}{R_\odot}\,.
\label{a_max}
\end{equation}
Here $\rho_\odot$ and $R_\odot$ are the density and the radius of the Sun, $\rho$ is the density of the asteroid, 
and $A_\mathrm{min}$ is the closest distance between the asteroid and the Sun.
Close to the Hill limit dynamics of the binary system gets much more complicated.
Moreover, the exact value of the Hill limit slightly depends on the mass fraction $\mu$.
Still, we neglect these fine details in our estimates.
The upper boundary of the shadowed area in Figure \ref{fig-plots} corresponds to $A_\mathrm{min}$=1 AU and $\rho$=2.5 g cm$^{-3}$.

We see that in many cases $a$ lies in the shadowed area.
These systems represent possible stably rotating doubly synchronous binaries.

\begin{figure}
 \centering
 \includegraphics[width=0.7\textwidth]{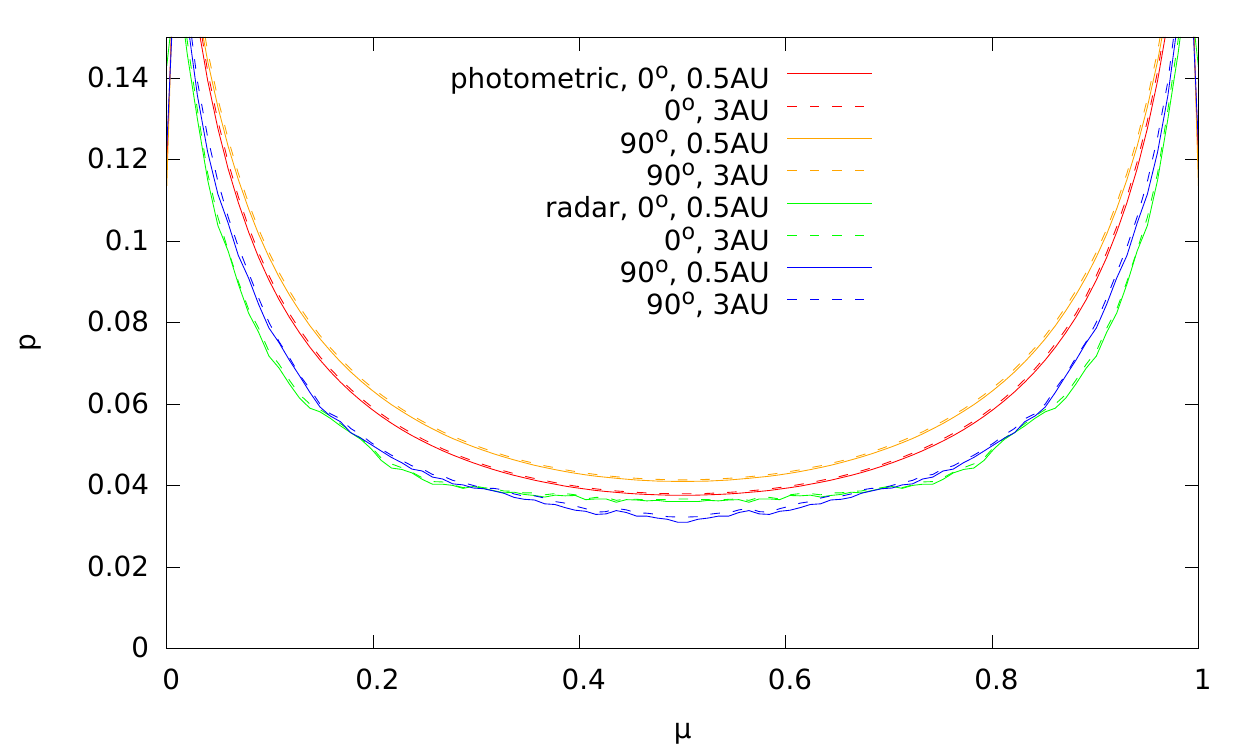}
 \caption{Dependence of probability of stable binary formation $p$ on the mass fraction $\mu$.}
 \label{fig-histogram}
\end{figure}

To estimate the probability of a doubly synchronous binary to end up in such a stable equilibrium for each value of $\mu$ evaluated,
we consider all possible asteroid shape pairs and orientations,
computing the total number of these that could reside in equilibrium state.
In Figure \ref{fig-histogram} such probabilities are ploted versus the mass fraction $\mu$
for two different obliquities $\varepsilon$, two different perihelion distances $A_\mathrm{min}$,
and two different sets of shape models.
We see that results for the both sets of models, both obliquities, and both perihelion distances are similar.
The probability $p$ of a binary ending up in a stable equilibrium is aboul 0.04 at $\mu \approx 0.5$,
and is over 0.1 for $\mu<0.1$ or $\mu>0.9$.
Binaries with very small or very large mass fractions are unlikely to become doubly synchronous,
but even the probability of 0.04 is significant enough to pay attention to,
as it could modify the persistence of a special class of doubly synchronous binaries in the asteroid population \citep{jacobson16}.

\section{Discussion}
\label{sec:discussion}

The common understanding of evolution of small asteroids is deduced from YORP cycles.
In this picture, each asteroid is accelerated by YORP to its disruption limit,
forms a binary, and the binary decays due to BYORP, tidal interaction, or chaotic dynamics.
Then the asteroid enters the next YORP cycle, and loses mass once again, and so on.
It has been argued by \cite{golubov12} that asteroids can stop their dynamical evolution 
because of stable equilibrium between tangential and normal YORP effects.

Here we find another possible equilibrium, this time on the binary stage of the YORP cycle. 
If each asteroid with some probability decays into a binary with such a shape and such a mass ratio that it appears in equilibrium,
than after sufficiently many YORP cycles such eqilibrium will necessary be attained.
Then the YORP evolution of the asteroid will be stopped for a very long time,
until a collision or a close encounter disrupts the otherwise stable system.
It implies that an asteroid can spend more time in stable equilibria than in YORP cycles,
that should drastically slow down the YORP evolution.

\label{lastpage}

\end{document}